\documentclass[12pt]{article}

\usepackage{epsfig}
\usepackage{array}
\usepackage{overcite}

\begin{document}

\newcommand{\pos}{\ensuremath{\mathbf{r}}}
\newcommand{\dt}{\ensuremath{\Delta t}}
\newcommand{\dr}{\ensuremath{\Delta \pos}}
\newcommand{\vi}{\ensuremath{\mathbf{v}_i}}
\newcommand{\vm}{\ensuremath{v}}
\renewcommand{\u}{\ensuremath{\mathbf{u}}}
\newcommand{\dab}{\ensuremath{\delta_{\alpha\beta}}}
\newcommand{\eq}[1]{equation (\ref{#1})}
\newcommand{\Eq}[1]{Equation (\ref{#1})}
\newcommand{\fig}[1]{fig. \ref{#1}}
\newcommand{\Fig}[1]{Fig. \ref{#1}}

\title{Jetting Micron-Scale Droplets onto Chemically Heterogeneous Surfaces}

\author{J. L\'{e}opold\`{e}s$^{1}$, A. Dupuis$^2$, D.G. Bucknall$^1$, J.M. Yeomans$^2$\\
\footnotesize{$^{1}$  Department of Materials, Oxford University, Oxford OX1 3PH, UK}\\
\footnotesize{$^{2}$  Department of Physics, Theoretical Physics, 1 Keble Road, Oxford OX1 3NP, UK}
}

\date{\today}

\maketitle

\begin{abstract}
  We report experiments investigating the behaviour of micron-scale
  fluid droplets jetted onto surfaces patterned with lyophobic and
  lyophilic stripes. The final droplet shape is shown to depend on the
  droplet size relative to that of the stripes. In particular when the
  droplet radius is of the same order as the stripe width, the final
  shape is determined by the dynamic evolution of the drop and shows a
  sensitive dependence on the initial droplet position and velocity.
  Lattice Boltzmann numerical solutions of the dynamical equations of
  motion of the drop provide a close quantitative match to the
  experimental results. This proves helpful in interpreting the data
  and allows for accurate prediction of fluid droplet behaviour for a
  wide range of surfaces.
\end{abstract}

\section{Introduction}

\label{sec:intro}

This paper presents a combined experimental and numerical
investigation of the behaviour of micron-scale fluid droplets jetted
onto chemically patterned surfaces~\cite{white1,Sirringhaus:00}. It
was motivated by questions of relevance to ink-jet printing where
substrates with chemical or physical defects can cause the expected
spherical shape of jetted droplets to become distorted, thus affecting
the integrity and quality of an image. However, the results have much
wider implications in the generic behaviour of fluid droplets on
heterogeneous surfaces. This is of particular interest since patterned
surfaces are used in wide diversity of areas such as electronic
devices, biologically active substrates or semiconductor
nanostructures.

Previous work on the behaviour of fluids on chemically patterned
substrates has predominantly concentrated on the equilibrium shape of
the drops~\cite{cass1}. In particular there has been extensive
theoretical and experimental work on the extent to which the line
tension affects droplet shape~\cite{whiteside1,white2,bosh1,bosh2}.
Several authors have considered the equilibrium configuration of a
fluid on one or two completely wetting stripes in a non-wetting
matrix~\cite{degennes2,shana1,china1,Darhuber:00,lipo2,Lipowsky:01,lipowski3,dietrich99}
and there is recent work describing how such a fluid moves along such
a stripe~\cite{str1}. By contrast, this paper describes the behaviour
of fluid droplets on chemically patterned surfaces when the fluid is
jetted onto the surface (non-zero impact velocity). This offers new
perspectives regarding the achievement of structures impossible to
obtain by simple droplet deposition.

The experimental results are compared to numerical simulations
achieved by a lattice Boltzmann solution of the equations of motion
for a one-component, two-phase fluid. Lattice Boltzmann models are a
class of numerical techniques ideally suited to probing the behaviour
of fluids on mesoscopic length scales~\cite{succi-book:01}. Several
lattice Boltzmann algorithms for a liquid-gas system have been
reported in the literature~\cite{swift:96,shan:93,he:98}. They solve
the Navier-Stokes equations of fluid flow but also input thermodynamic
information, typically either as a free energy or as effective
microscopic interactions. They have proved successful in modelling
such diverse problems as fluid flows in complex
geometries~\cite{succi:89}, two-phase models~\cite{swift:96,shan:93},
hydrodynamic phase ordering~\cite{kendon:99} and sediment transport in
a fluid~\cite{dupuis:00b}.

We consider a one-component, two-phase fluid and use the free energy
model originally described by Swift et al.~\cite{swift:96} with a
correction to ensure Galilean invariance~\cite{holdych:98}. The
advantage of this approach for the wetting problem is that it allows
us to tune equilibrium thermodynamic properties such as the surface
tension or static contact angle to agree with analytic predictions.
Thus it is rather easy to control the wetting properties of the
substrate.

Choosing fluid parameters such as viscosity, surface tension and
contact angles to match the experimental values we obtain a close
quantitative match to the different droplet shapes thus helping to
understand the mechanisms for their formation. The results show the
importance of dynamic and metastability effects in determining the
shapes formed by small droplets on heterogeneous substrates. In
particular we demonstrate that the final droplet configuration depends
on the droplet size relative to that of the stripes, the initial point
of impact and the incident droplet velocity.

\section{Experimental Section}

\subsection{Fluid}

The solventless UV cure black ink-jet ink used (acrylate monomer) has
a viscosity of 25.0 mPa.s and a surface tension of 24.4 mN.m$^{-1}$.
The droplets were jetted using a $256$ nozzle industrial inkjet
printhead held $1$ mm from the substrate. The print pattern was
configured so that the droplets impacted on the surface with an
average separation between each drop of 280 $\mu$m, at a velocity of 8
m.s$^{-1}$. The ink droplets were cured $1.6$ s after ejection from
the print head using a standard mercury H UVA lamp system. The radius
of the droplets before impact was chosen to be $22$ $\mu$m, and the
typical spreading time of the droplets is of order milliseconds.

\subsection{Preparation of the substrates}

Surfaces are produced with areas of different wettabilities using
standard microcontact printing techniques. This is nowadays a well
established method, and abundant literature can be found on the
subject~\cite{white1}. The chemical patterns are created on gold
coated silicon $(001)$ wafers to make certain that the substrates are
molecularly flat, thereby ensuring any effects observed in the droplet
behaviour resulted from chemical rather topographic (or a combination
of both) effects. Prior to gold evaporation, the wafers are coated
with 50 nm chromium to ensure stability of the gold (100 nm). The
samples are washed with ethanol and dried with a stream of $N_{2}$
before patterning.

\begin{table}
\caption{Widths of the chemically patterned stripes used in this study, 
as determined by scanning electron microscopy. The sample numbering
corresponds to that used in Figure~\ref{topview2}.}
\label{sizstripes}
\begin{center} 
\begin{tabular}{ccc}  
Sample & lyophilic ($\mu$m) & lyophobic ($\mu$m)\\        
\hline        
S1&$5$&$56$\\        
S2&$13$&$65$\\        
S3&$19$&$35$\\        
S4&$26$&$47$\\        
S5&$23$&$38$\\        
S6&$50$&$50$\\        
S7&$47$&$31$\\                 
S8&$83$&$75$\\
\end{tabular}
\end{center}
\end{table}

The stripes are created using solutions of 4 mmol of either methyl- or
carboxyl- terminated octadecyl thiols, which produced lyophobic
(--CH$_{3}$) and lyophilic (--COOH) monolayer regions on the Au
surface.  To achieve this, a polydimethylsiloxane stamp is moulded on
a master having the desired pattern, and allowed to cure for one week
at room temperature. Then a known quantity of hexadecanethiol solution
(4 mmol in hexane) is poured onto the elastomeric stamp. After 10
seconds and drying with a stream of $N_{2}$, the stamp is applied onto
the surface of the wafer.

This produces a lyophobic self assembled monolayer (SAM) with the
desired pattern. The sample is then immerged one hour in a mercapto
undecanoic acid / ethanol solution (4mmol) in order to complete the
patterning.

The corresponding contact angles for the commercially available black
UV cure jet-ink used in these experiments are respectively
64$^{\circ}$ and 5$^{\circ}$ on the lyophobic and lyophilic areas. The
contact angles are determined by optical measurement of the
equilibrium shape of equivalent sized droplets on surfaces
homogeneously produced by solution casting the relevant thiol. The
widths of the stripes created by the microcontact printing process,
and measured using scanning electron microscopy, are listed in Table
\ref{sizstripes}.

\section{The lattice Boltzmann model}

\label{sec:model}

The lattice Boltzmann approach solves the Navier-Stokes equations by
following the evolution of partial distribution functions $f_i$ on a
regular, $d$-dimensional lattice formed of sites $\pos$. The label $i$
denotes velocity directions and runs between $0$ and $z$. $DdQz+1$ is
a standard lattice topology classification. The $D3Q15$ lattice
topology we use here has the following velocity vectors $\vi$:
$(0,0,0)$, $(\pm 1,\pm 1,\pm 1)$, $(\pm 1,0,0)$, $(0,\pm 1, 0)$,
$(0,0, \pm 1)$ in lattice units.

The lattice Boltzmann dynamics are given by
\begin{equation}
f_i(\pos+ \dt \vi,t+\dt)=f_i(\pos,t)+\frac{1}{\tau}\left(f_i^{eq}(\pos,t)-f_i(\pos,t)\right)
\label{eq:lbDynamics}
\end{equation}
where $\dt$ is the time step of the simulation, $\tau$ the relaxation
time and $f_i^{eq}$ the equilibrium distribution function which is a
function of the density $n=\sum_{i=0}^z f_i$ and the fluid velocity
$\u$ defined through the relation $n\u=\sum_{i=0}^z f_i\vi$.

The relaxation time tunes the kinematic viscosity as~\cite{succi-book:01}
\begin{equation}
\nu=\frac{\dr^2}{\dt} \frac{C_4}{C_2} (\tau-\frac{1}{2})
\label{eq:visco}
\end{equation}
where $\dr$ is the lattice spacing and $C_2$ and $C_4$ are
coefficients related to the topology of the lattice. These are equal
to $3$ and $1$ respectively when one considers a $D3Q15$ lattice
(see~\cite{dupuis:02} for more details).

It can be shown~\cite{swift:96} that equation~(\ref{eq:lbDynamics})
reproduces the Navier-Stokes equations of a non-ideal gas if the local
equilibrium functions are chosen as
\begin{eqnarray}
f_i^{eq}&=&A_\sigma+B_\sigma u_\alpha v_{i\alpha} + C_\sigma \u^2
         +D_\sigma u_\alpha u_\beta v_{i\alpha}v_{i\beta} 
         +G_{\sigma\alpha\beta} v_{i\alpha}v_{i\beta}, \quad i>0,
         \nonumber \\
f_0^{eq}&=& n - \sum_{i=1}^z f_i^{eq}
\label{eq:lbEq}
\end{eqnarray}
where Einstein notation is understood for the Cartesian labels
$\alpha$ and $\beta$ (i.e.  $v_{i\alpha}u_\alpha=\sum_\alpha
v_{i\alpha}u_\alpha$) and where $\sigma$ labels velocities of
different magnitude. A possible choice of the coefficients
is~\cite{dupuis:03}
\begin{eqnarray}
A_\sigma & = & \frac{w_\sigma}{c^2}\left( p_b- 
               \frac{\kappa}{2} (\partial_\alpha n)^2               
               -\kappa n \partial_{\alpha\alpha} n 
               + \nu u_\alpha \partial_\alpha n \right), \nonumber \\
B_\sigma & = & \frac{w_\sigma n}{c^2}, \quad 
  C_\sigma = -\frac{w_\sigma n}{2 c^2}, \quad 
  D_\sigma = \frac{3 w_\sigma n}{2 c^4}, \nonumber \\
G_{1\gamma\gamma} & = & \frac{1}{2 c^4} \left( \kappa(\partial_\gamma n)^2 +2 
\nu u_\gamma \partial_\gamma n \right) , \quad 
  G_{2\gamma\gamma} = 0, \nonumber \\
G_{2\gamma\delta} & = & \frac{1}{16 c^4} \left( \kappa (\partial_\gamma n)
  (\partial_\delta n) + \nu (u_\gamma \partial_\delta n + u_\delta
 \partial_\gamma n) \right)
\label{lb:eqCoeff}
\end{eqnarray}
where $w_1=1/3$, $w_2=1/24$, $c=\dr/\dt$, $\kappa$ is a parameter
related to the surface tension and
$p_b=p_c(\nu_p+1)^2(3\nu_p^2-2\nu_p+1-2\beta\tau_p)$ is the pressure
in the bulk where $\nu_p=(n-n_c)/n_c$, $\tau_p=(T_c-T)/T_c$ and
$p_c=1/8$, $n_c=3.5$ and $T_c=4/7$ are the critical pressure, density
and temperature respectively and $\beta$ is a constant typically equal
to $0.1$.

The derivatives in the direction normal to the substrate are handled
in such a way that the wetting properties of the substrate can be
controlled. A boundary condition can be established using the Cahn
model~\cite{cahn:77}. He proposed adding an additional surface free
energy $\Psi_c(n_s)=\phi_0-\phi_1 n_s+\cdots$ at the solid surface
where $n_s$ is the density at the surface. Neglecting the second order
terms in $\Psi_c(n)$ and minimizing $\Psi_b+\Psi_c$ (where $\Psi_b$ is
the free energy in the bulk), a boundary condition valid at $z=0$
emerges
\begin{equation}
\partial_z n= - \frac{\phi_1}{\kappa}.
\label{eq:cahn1}
\end{equation}

\Eq{eq:cahn1} is imposed on the substrate sites to implement the Cahn
model in the lattice Boltzmann approach. Details are given
in~\cite{briant:02}.

The Cahn model can be used to relate $\phi_1$ to $\theta$ the contact
angle defined as the angle between the tangent plane to the droplet
and the substrate~\cite{dupuis:03}
\begin{equation}
\phi_1=2 \beta \tau_p \sqrt{2p_c \kappa} \; \mathrm{sign}(\theta -
\frac{\pi}{2}) \sqrt{\cos\frac{\alpha}{3}\left(1-\cos\frac{\alpha}{3}\right)}
\label{eq:phi1}
\end{equation}
where $\alpha=\mathrm{cos}^{-1}(\sin^2\theta)$ and the function
$\mathrm{sign}$ returns the sign of its argument.

We impose a no-slip boundary condition on the velocity. Because the
full dynamics takes place on the boundary the usual bounce-back
condition must be extended to ensure mass conservation (see
\cite{dupuis:02} for a wider discussion). This is done by a suitable
choice of the rest field, $f_0$, to correctly balance the mass of the
system.

This model reproduces Young's law and the expected dependence of the
droplet behaviour on viscosity and surface tension~\cite{dupuis:03}.

The following lattice Boltzmann parameters are set. The initial
droplet radius $R_0=30$ lattice sites. The droplet is initialised with
a vertical velocity equal to $U_0=0.02$. The lattice geometry is $L_x
\times L_y \times L_z$ where $L_x$ and $L_y$ are chosen large enough
to not affect the behaviour of the droplet and $L_z=40$. The
relaxation time $\tau=0.63$. The surface tension related parameter
$\kappa=0.0012$.  The temperature $T=0.4$ which leads to two phases of
density $n_l=4.128$ and $n_g=2.913$. The simulations are run for
$400~000$ iterations.

Simulation and physical parameters are related as usual by choosing a
length scale $L_0$, a time scale $T_0$ and a mass scale $M_0$. A
simulation parameter with dimensions $[L]^{n_1}[T]^{n_2}[M]^{n_3}$ is
multiplied by $L_0^{n_1}T_0^{n_2}M_0^{n_3}$ to give the physical
value. $L_0=1.5 \cdot 10^{-6} m$, $T_0=3.7 \cdot 10^{-9} s$ and
$M_0=7.9 \cdot 10^{-16} kg$ were chosen to give physically realistic
droplet diameter, viscosity and surface tension.

\section{Results and discussion}

\begin{figure}[htbp] 
\begin{center}
\epsfig{file=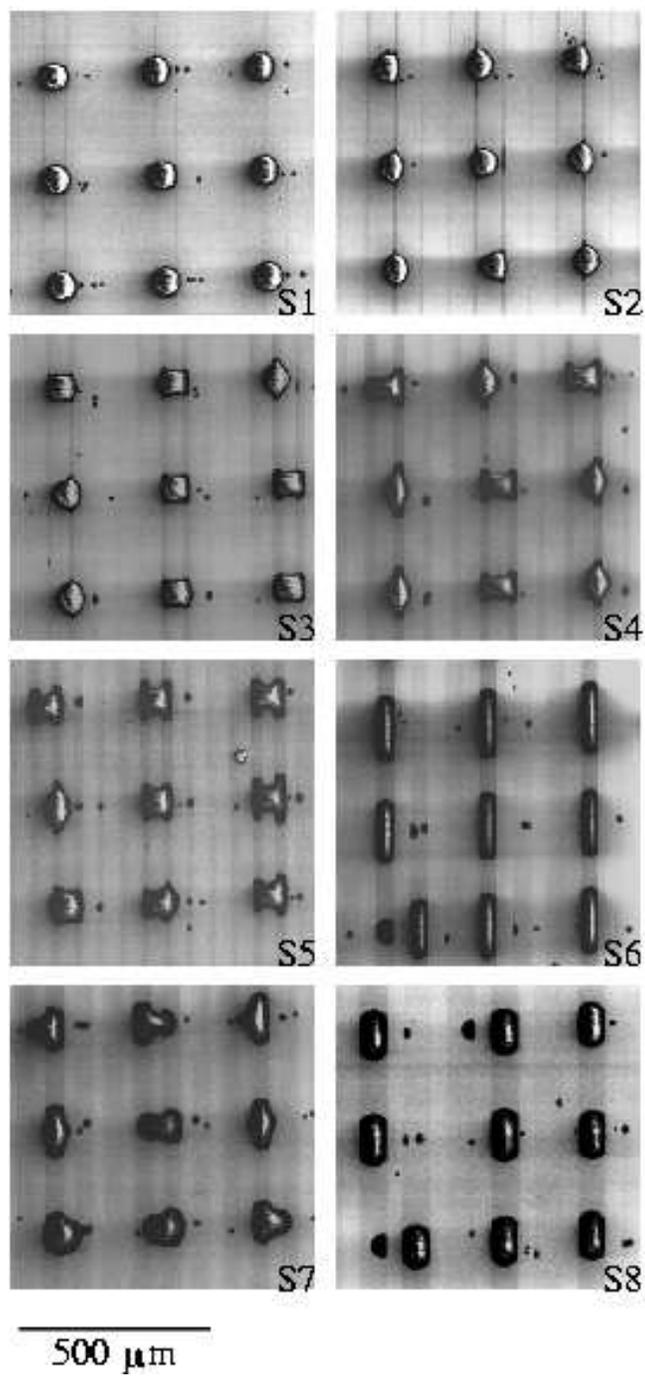,width=8.5cm}
\end{center} 
\caption{Scanning electron micrographs of inkjet droplets on patterned   
  surfaces.  Lyophilic and lyophobic stripes appear dark and pale
  respectively \protect \cite{japo1}. The stripe widths for the
  different samples are listed in Table~\ref{sizstripes}.}
\label{topview2}
\end{figure}

A top view of the final droplet shapes on the different patterned
surfaces is shown in Figure \ref{topview2}. To our knowledge, it is
the first time that the evolution of droplet shape has been studied
over such a large range of heterogeneities.

When the width of the stripes is much smaller than the droplet radius
($S1$) the contact line is almost circular with only small deviations
near the stripe boundaries as previously reported in the
literature~\cite{whiteside1,white2}.

For wider stripes, where the dimensions are now approaching the
droplet diameter, the fluid remains confined to a lyophilic region
($S6,S8$) and the final shape is therefore highly elongated.

It should also be noted that in samples $S6$ and $S8$, the regular
droplet pattern array is disrupted as droplets can be displaced to the
neighbouring lyophilic region. When this occurs, a small portion of
the initial droplet volume can remain on a neighbouring lyophilic
stripe.

In the intermediate cases, for droplets impinging on the stripes $S3$,
$S4$, $S5$ and $S7$, two different characteristic droplet shapes are
observed: "lozenge" or "butterfly".

The reason for the formation of these shapes can be understood by
considering a lattice Boltzmann modelling of the droplet dynamics.

\begin{figure}
\begin{center}
\begin{tabular}{c}
(a) \\
\vspace{0mm}\\
\epsfig{file=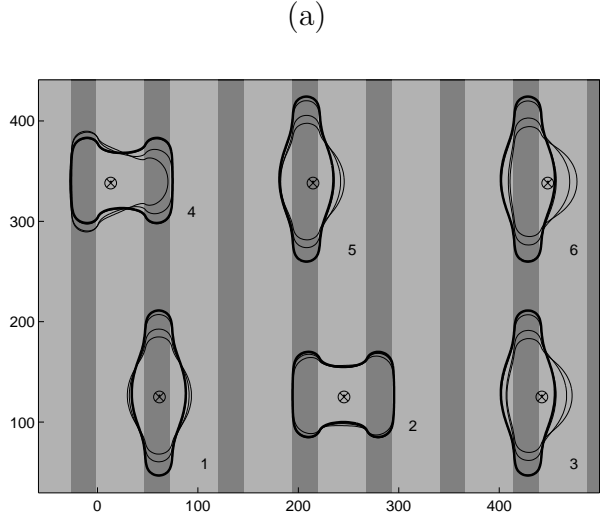,width=7.9cm} \\
\vspace{2mm}\\
(b) \\
\vspace{0mm}\\
\epsfig{file=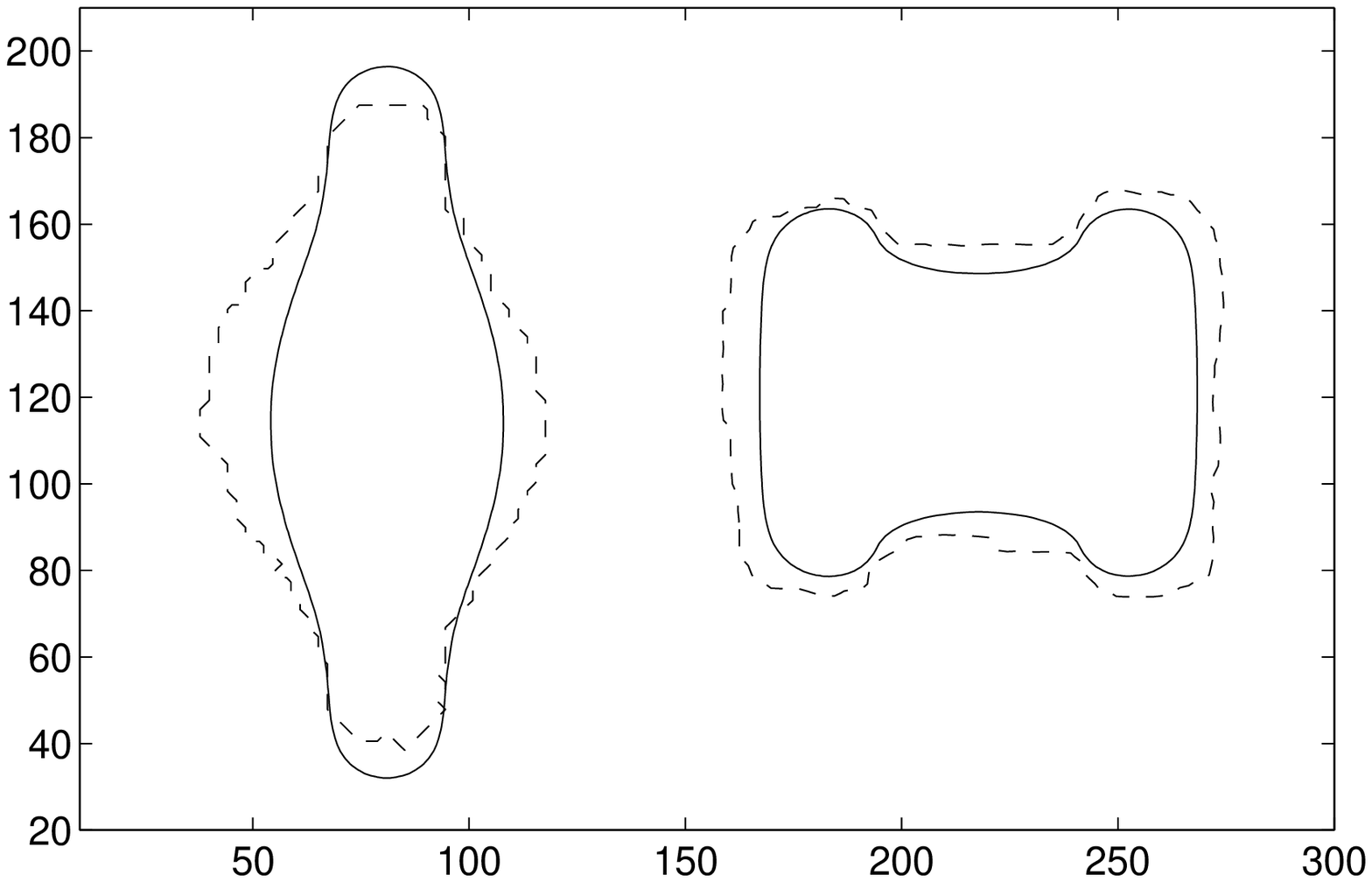,width=6cm} \\
\end{tabular}
\end{center}
\caption{The modelled substrate is defined so that the surface 
  heterogeneities are equivalent with the sample $S4$. (a) Numerical
  simulation of droplets hitting the surface at various impact points
  indicated by encircled crosses. For each droplet the bold and faint
  lines represent the extent of the droplet at equilibrium and at
  intermediate times, respectively. The lyophobic and lyophilic
  regions are shaded to be consistent with Figure \ref{topview2}. (b)
  Direct comparison between experimental ($S4$, dashed lines) and
  numerical (solid lines) equilibrium contact lines. The length scales
  in both plots are reported in micrometers.}
\label{drops}
\end{figure}

Figure \ref{drops}(a) shows the droplet contact lines for intermediate
and equilibrium stages of the droplet wetting as a function of the
initial impact position of the jetted fluid. The modelled substrate is
defined so that the surface heterogeneities are equivalent with the
sample $S4$. The droplet at the point of impact is assumed to be
spherical with an imposed velocity equal to that used experimentally
($8$ m.s$^{-1}$). It is quite clear that the shapes obtained by such
calculations are entirely consistent with those obtained
experimentally. The very strong similarities are more clearly
demonstrated in Figure \ref{drops}(b) where the experimental and
numerical simulations are overlaid for the two equilibrium droplet
shapes. The good quantitative agreement between the simulations and
the experimental results is extremely pleasing given that all the
parameters in the simulation are fixed by the physical conditions
dictated by the experiments. Small differences are not surprising
because of the uncertainty in the experimental values of the transport
coefficients and surface energies, and the possibility of contact line
pinning on real substrates.

Note in particular that, just as in the experiments, two different
final droplet shapes are obtained in the numerical modelling for the
substrate geometry $S4$.

The simulations allow us to follow the dynamical pathway by which the
final equilibrium states are reached. Consider case 1 (see
Figure~\ref{drops}(a)) where the initial contact point of the droplet
is in the centre of a lyophilic stripe. The drop initially evolves
symmetrically parallel and normal to the stripe. Once the boundary of
the lyophobic stripe is reached horizontally, the fluid moves more
quickly along the lyophilic than along the lyophobic part of the
substrate. Note that there is an overshoot of the fluid in the
direction perpendicular to the stripes (the maximum extent of the
contact line exceeds that of the equilibrium droplet). This occurs as
the droplet initially spreads due to its impact velocity, and relaxes
to its equilibrium shape determined by the wetting behaviour. The net
effect is an equilibrium shape which has a distinctive lozenge shape
centred over a lyophilic stripe.

Now consider case $2$ where the droplet strikes the substrate in the
middle of a lyophobic stripe. Because the initial diameter is
approximately the same as that of the lyophobic stripes, the droplet
spreads directly onto adjacent lyophilic parts of the substrate giving
the characteristic butterfly shape which is symmetrically located over
two lyophilic and one lyphobic stripes.

When the droplet impacts on a location other than the middle of a
stripe, the symmetry of the evolution parallel to the stripes is lost.
Despite this initial asymmetry the final state is always symmetric and
produces either the characteristic lozenge (for the initial positions
$3$, $5$ and $6$) or butterfly (for the initial position $4$)
patterns. These states correspond to long lived metastable or stable
equilibria~\cite{lipowski3}. The numerical results give a free energy
for the butterfly-shaped droplets which is $\sim 1 \%$ of the lozenge
free energy.

Figure~\ref{drops}(a) also allows us to predict that jetting droplets
onto substrate $S4$ should lead to approximately equal numbers of
lozenge and butterfly shapes. Numerical results show that a droplet
spreads into a lozenge if it hits the surface on a lyophilic stripe or
its close surroundings (up to about one eight of the width of the
neighbouring lyophobic stripes). This corresponds to about half the
substrate area and therefore one expects about one half of the
droplets to take each of the final shapes consistent with the
experimental results for $S4$, as shown in

Figure~\ref{topview2}. Thus far we have shown that the parameters
affecting the final shape of a drop impacting on a striped substrate
are the relative sizes of the drop and the stripes and the initial
point of impact.

\begin{figure}
\begin{center}
\begin{tabular}{m{1.5cm}m{3cm}m{0.3cm}m{3cm}}
 & \centerline{ \hspace*{0.2cm}(a)} & & \centerline{\hspace*{0.2cm}(b)} \\
\centerline{0.03 $t_{f}$}  & \epsfig{file=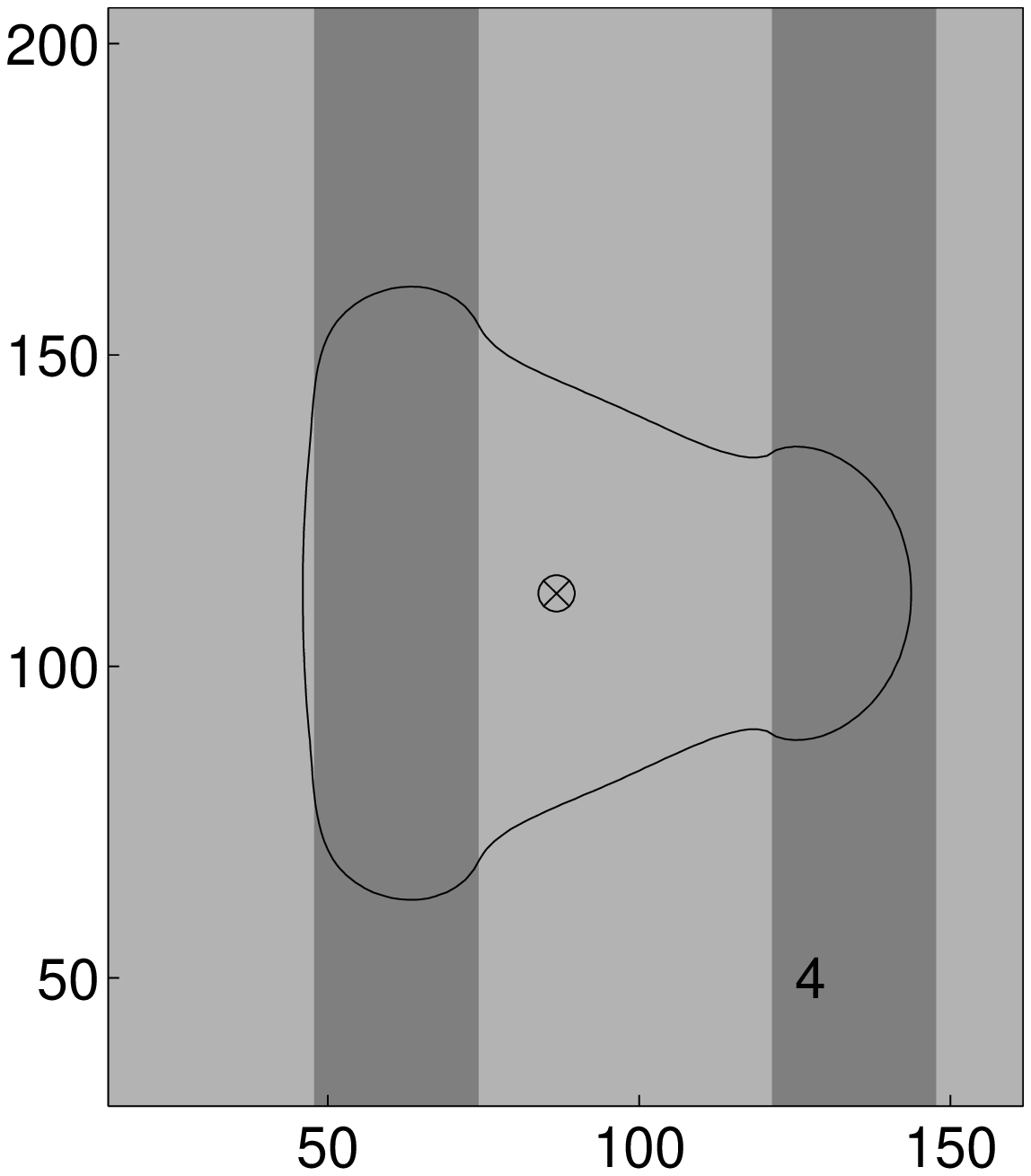,width=30mm} & &
                             \epsfig{file=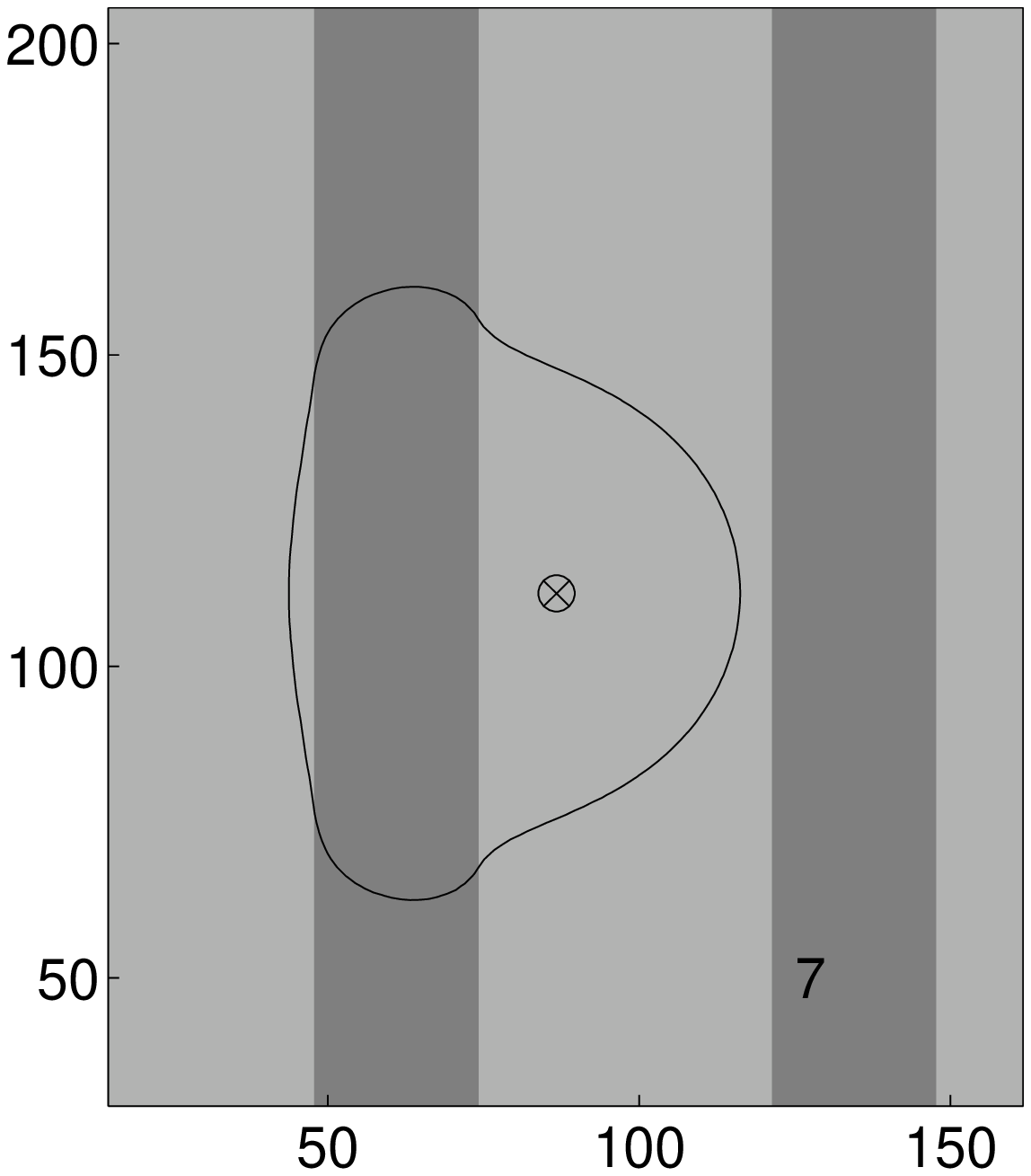,width=30mm}  \\ 
 \centerline{0.05 $t_{f}$} & \epsfig{file=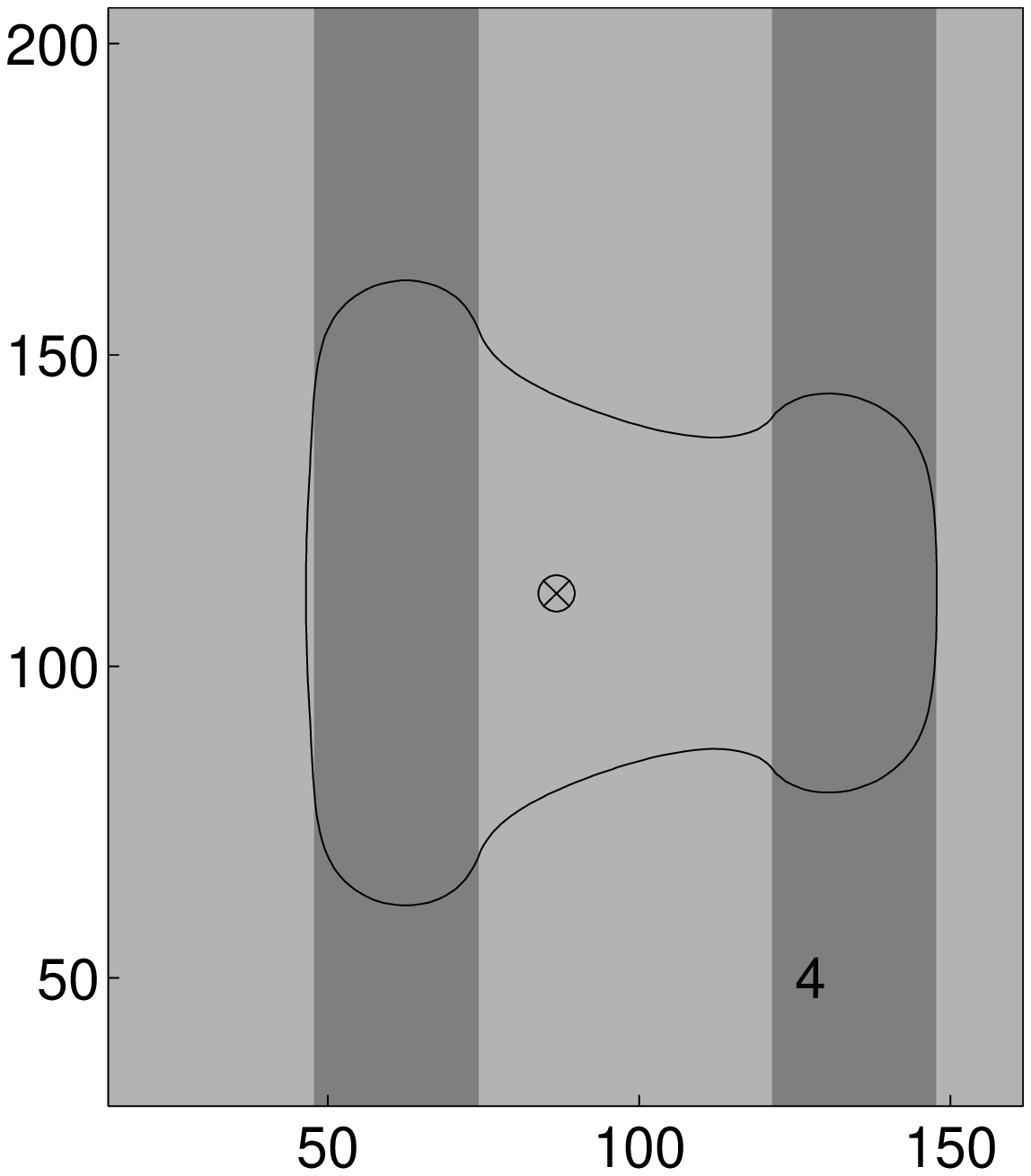,width=30mm} & &
                             \epsfig{file=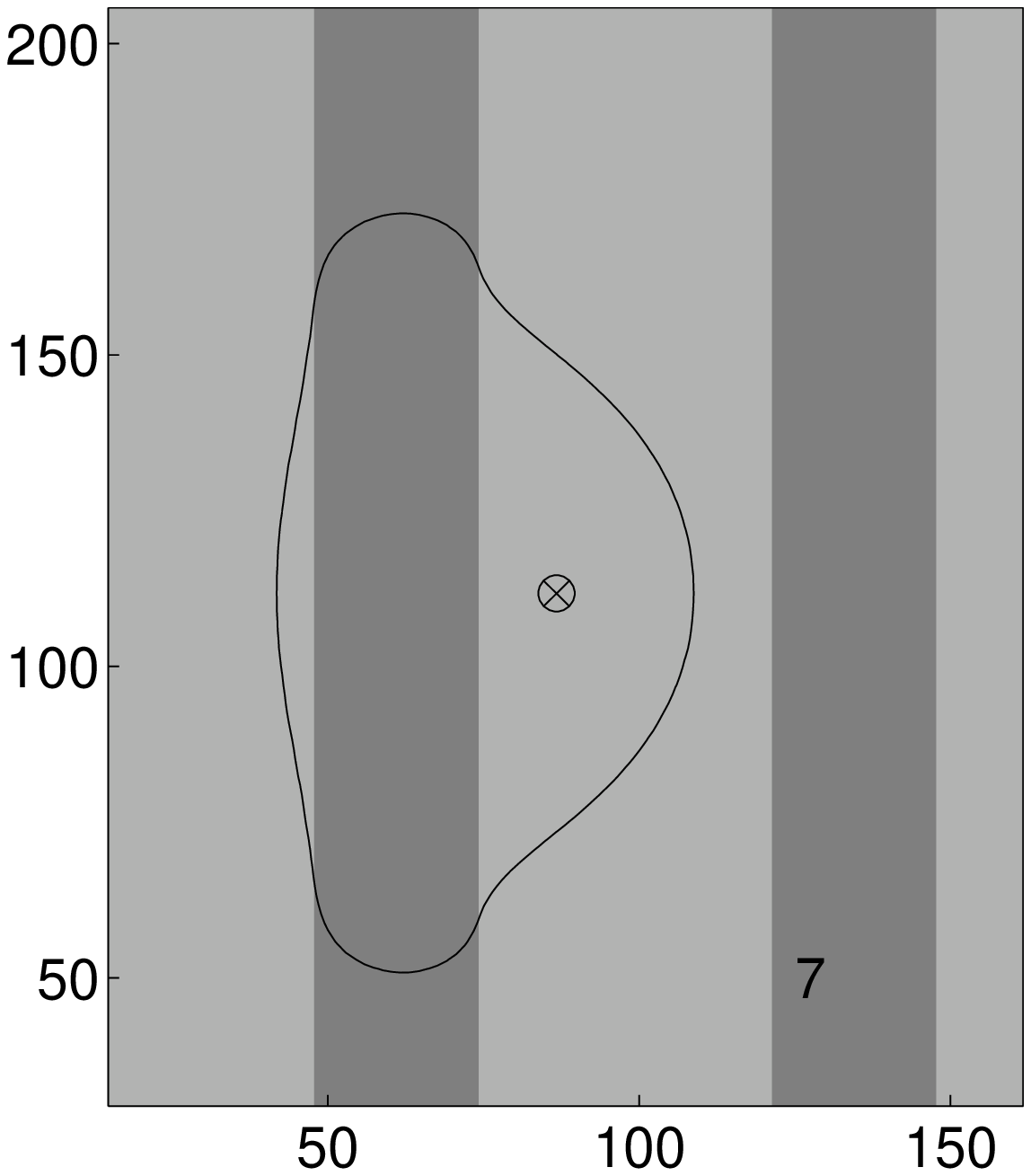,width=30mm}  \\ 
 \centerline{0.3 $t_{f}$}  & \epsfig{file=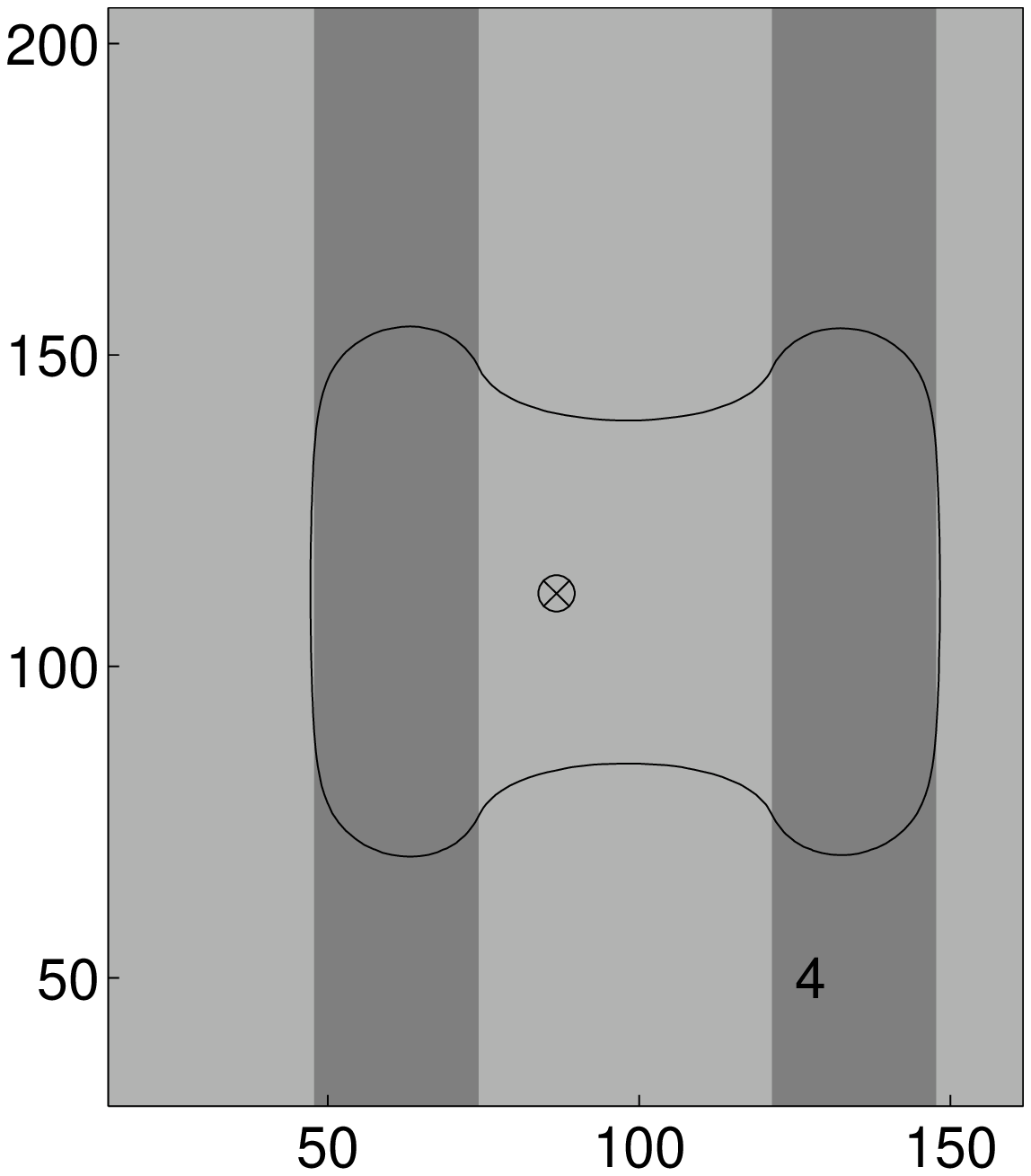,width=30mm} & &
                             \epsfig{file=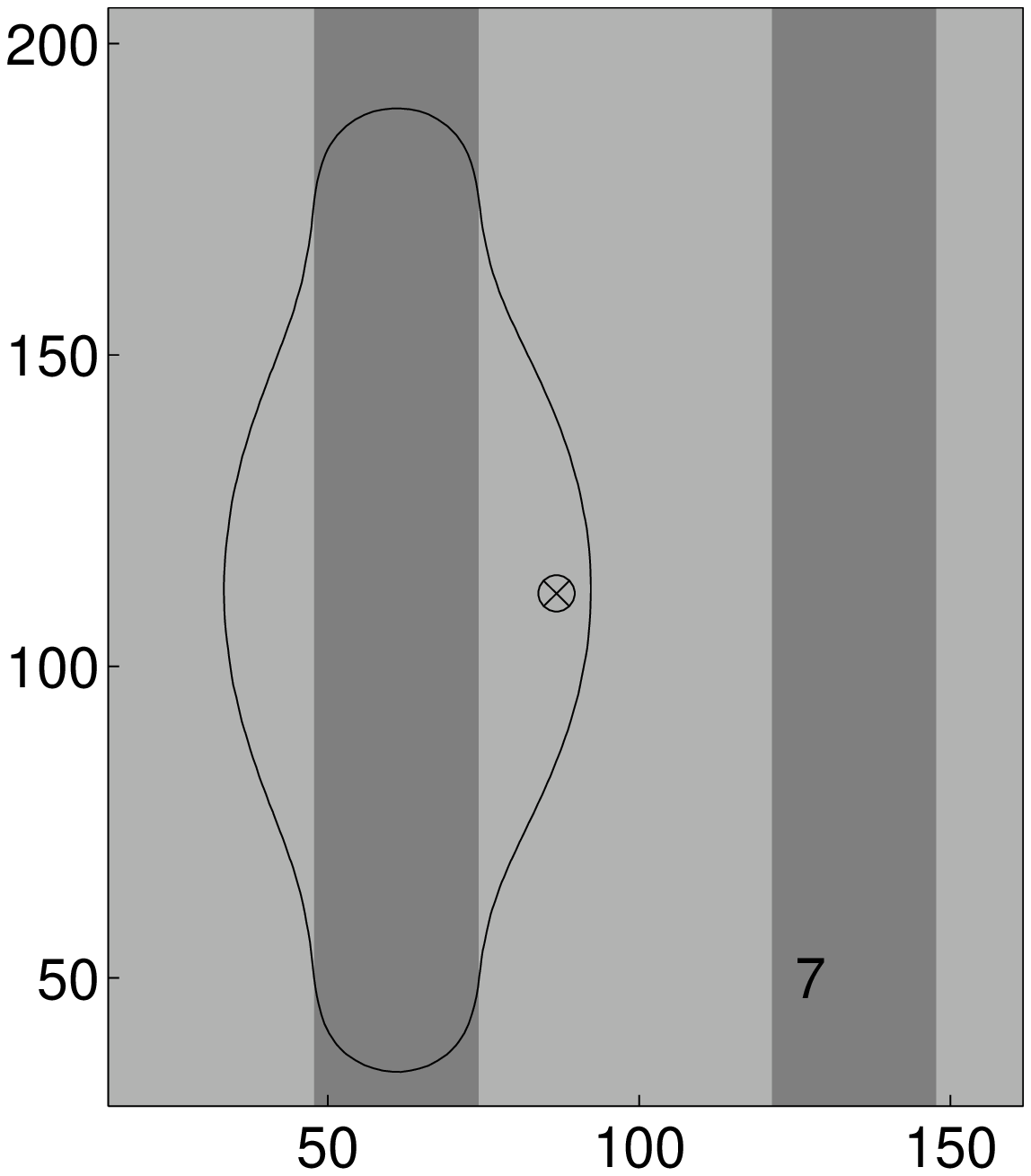,width=30mm}  \\ 
 \centerline{$t_{f}$}      & \epsfig{file=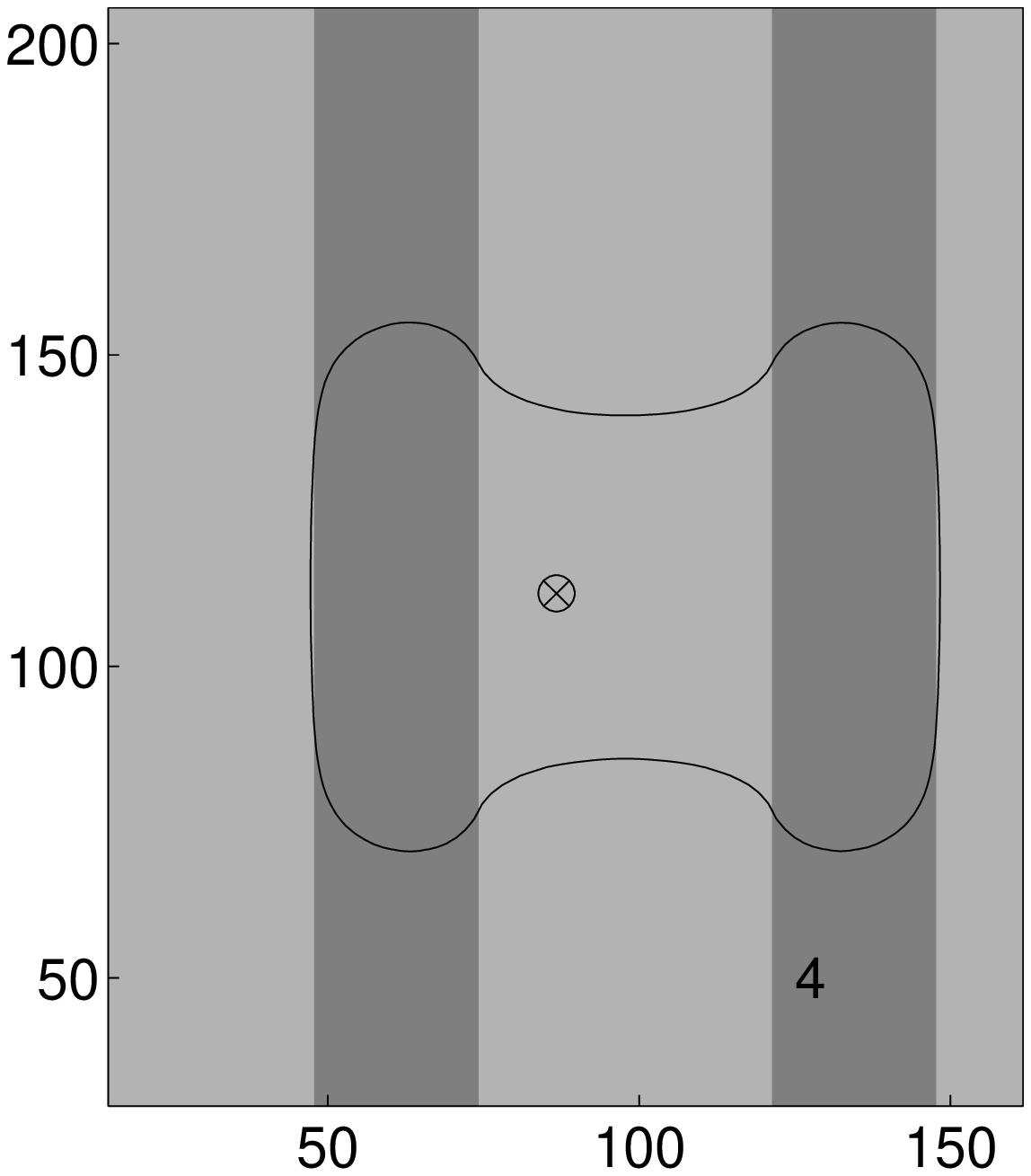,width=30mm} & &
                             \epsfig{file=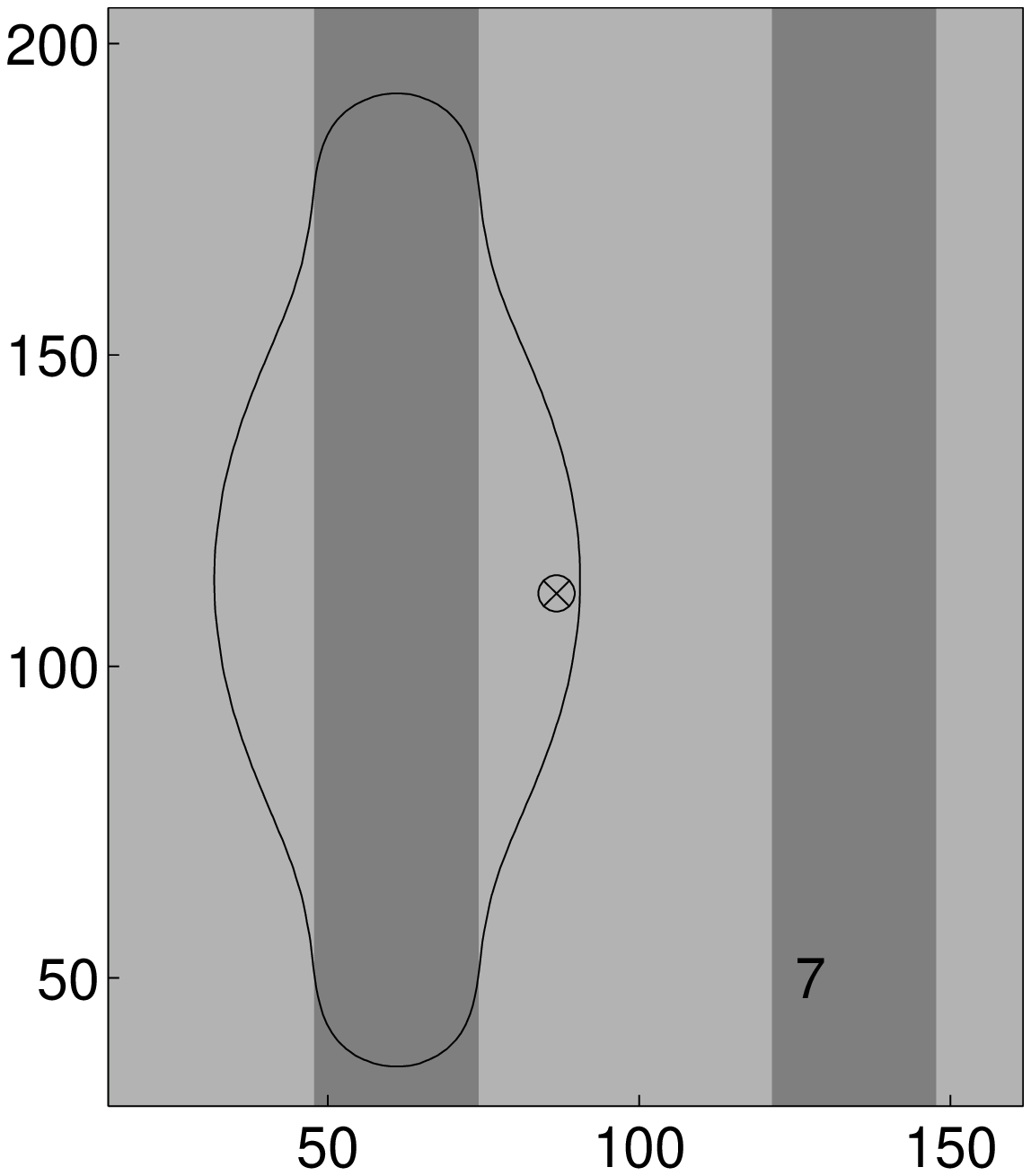,width=30mm}  
\end{tabular}
\end{center}
\caption{Time evolution (top to bottom) of the droplet shape from 
  numerical simulations of the equations of motion. (a) Droplet
  spreading with impact velocity 8 ms$^{-1}$. (b) No impact velocity.
  The initial point of impact is the same for both simulations.
  $t_{f}$ is the final simulation time.}
\label{dropwithAndwithoutVel}
\end{figure}

Clearly, the surface energies of the different stripes will have an
important effect on the droplet spreading behaviour~\cite{lipowski3},
and given the results described above the initial drop velocity may
also be important.

To test the latter, numerical simulations comparing the time evolution
of a drop with and without impact velocity have been undertaken (see
Figure~\ref{dropwithAndwithoutVel}). The final state is indeed
different, as the drop with no impact velocity is unable to reach, and
hence take advantage of the wetting possibilities of the neighbouring
lyophilic regions of the substrate. For a given average distance
between heterogeneities, the droplet velocity appears to be an
important parameter regarding to the number of surface heterogeneities
encountered during spreading.

\section{Conclusion}

We have presented experimental and numerical results investigating the
behaviour of micron-scale droplets on chemically patterned substrates.
When the drop radius is of the same order as the stripe width the
final droplet shape is determined by the dynamic evolution of the drop
and is very sensitive to the initial droplet position and velocity.
The final state may be metastable and is not determined just by a
minimisation of the free energy.

We have shown that it is possible to provide a close quantitative
correspondence between numerical solutions of the hydrodynamic
equations of motion describing the spreading and the experimental
results. This has proved invaluable in fully understanding the data
and in predicting droplet behaviour for parameter values not available
experimentally.

These results underline the difficulties inherent in controlling the
details of patterns formed using ink-jet printing (where the
underlying substrate is likely to have both chemical and topological
heterogeneities) and the subtle effects of the surface wetting
properties on the behaviour of liquids on patterned substrates. The
present work will therefore be extended to topological surface
features and complex fluids.

\section*{Acknowledgments}

We gratefully thank Dr. G. Desie and S. Allaman (Agfa Research Center,
Belgium) for their collaboration. Dr. H. Zhang is acknowledged for
allowing us to use the PDMS stamp. This work forms part of the
IMAGE-IN project which is funded by the European Community through a
Framework 5 grant GRD1-CT-2002-00663.

\bibliographystyle{unsrt}

\begin{thebibliography}{10}

\bibitem{white1}
Y.~Xia and G.~M. Whitesides.
\newblock {\em Angew. Chem. Int.}, 37:550--575, 1998.

\bibitem{Sirringhaus:00}
H.~Sirringhaus, T.~Kawase, R.H. Friend, T.~Shimoda, and M.~Inbasekaran.
\newblock {\em Science}, 290:2123--2126, 2000.

\bibitem{cass1}
A.~B.~D. Cassie.
\newblock {\em Discuss. Faraday Soc.}, 3:11, 1948.

\bibitem{whiteside1}
J.~Drelich, J.~L. Wilbur, J.~D. Miller, and G.~M. Whitesides.
\newblock {\em Langmuir}, 12(7):1913--1922, 1996.

\bibitem{white2}
J.~Drelich, J.~D. Miller, A.~Kumar, and G.~M. Whitesides.
\newblock {\em Coll. Surf., A: Physicochem. Eng. Aspects}, 93:1, 1994.

\bibitem{bosh1}
T.~Pompe, A.~Fery, and S.~Herminghaus.
\newblock {\em Langmuir}, 14(10):2585--2588, 1998.

\bibitem{bosh2}
J.~Buerhle, S.~Herminghaus, and F.~Mugele.
\newblock {\em Langmuir}, 18:9771--9777, 2002.

\bibitem{degennes2}
P.~G. de~Gennes and J.~F. Joanny.
\newblock {\em J. Chem. Phys.}, 81(1):552--562, July 1984.

\bibitem{shana1}
M.~E.~R. Shanahan.
\newblock {\em Coll. and Surf. A: Physichochem. Eng. Aspects.}, 156:71--77,
  1999.

\bibitem{china1}
D.~Li.
\newblock {\em Coll. and Surf. A: Physicochem. Eng. Aspects}, 116:1--23, 1996.

\bibitem{Darhuber:00}
A.~A. Darhuber, S.~M.Troian, and S.~M. Miller.
\newblock {\em J. Appl. Phys.}, 87(11):7768--7775, 2000.

\bibitem{lipo2}
P.~Lenz.
\newblock {\em Adv. Mater.}, 11(18):1531--1534, 1998.

\bibitem{Lipowsky:01}
R.~Lipowsky.
\newblock {\em Curr. Opin. Colloid Interface Sci.}, 6:40--48, 2001.

\bibitem{lipowski3}
M.~Brinkmann and R.~Lipowsky.
\newblock {\em J. Appl. Phys.}, 92(8):4296--4306, 2002.

\bibitem{dietrich99}
C.~Bauer and S.Dietrich.
\newblock {\em Phys. Rev. E}, 60(6):6919--6941, 1999.

\bibitem{str1}
A.~A. Darhuber, S.~M. Troian, and W.~W. Reisner.
\newblock {\em Phys. Rev. E}, 64:031603, 2001.

\bibitem{succi-book:01}
S.~Succi.
\newblock {\em The Lattice {B}oltzmann Equation, For Fluid Dynamics and
  Beyond}.
\newblock Oxford University Press, 2001.

\bibitem{swift:96}
M.R. Swift, E.~Orlandini, W.R. Osborn, and J.M. Yeomans.
\newblock Lattice {B}oltzmann simulations of liquid-gas and binary fluid
  systems.
\newblock {\em Phys. Rev. E}, 54:5051--5052, 1996.

\bibitem{shan:93}
X.~Shan and H.~Chen.
\newblock {\em Phys. Rev. E}, 47:1815--1819, 1993.

\bibitem{he:98}
X.~He, S.~Chen, and G.D. Doolen.
\newblock {\em J. Comput. Phys.}, 146:282--300, 1998.

\bibitem{succi:89}
F.~Higuera, S.~Succi, and E.~Foti.
\newblock {\em EuroPhysics Letters}, 10(5):433--438, 1989.

\bibitem{kendon:99}
V.M. Kendon, J.C. Desplat, P.~Bladon, and M.E. Cates.
\newblock {\em Phys. Rev. Lett.}, 83(3):576--579, 1999.

\bibitem{dupuis:00b}
A.~Dupuis and B.~Chopard.
\newblock {\em J. Comput. Phys.}, 178(1):161--174, 2002.

\bibitem{holdych:98}
D.~Holdych, D.Rovas, J.~Georgiadis, and R.~Buckius.
\newblock {\em Int. {J}. {M}od. {P}hys. {C}}, 9:1393--1404, 1998.

\bibitem{dupuis:02}
A.~Dupuis.
\newblock {\em From a lattice {B}oltzmann model to a parallel and reusable}.
\newblock PhD thesis, University of Geneva,
  http://cui.unige.ch/spc/PhDs/aDupuisPhD/phd.html, June 2002.

\bibitem{dupuis:03}
A.~Dupuis, A.~J. Briant, C.~M. Pooley, and J.~M. Yeomans.
\newblock In P.M.A~Sloot et~al., editor, {\em Proceedings of the {ICCS 2003}
  conference}, pages 1024--1033, St.Petersburg, Russia, 2003. Springer.

\bibitem{cahn:77}
J.~W. Cahn.
\newblock Critical point wetting.
\newblock {\em J. Chem. Phys.}, 66:3667--3672, 1977.

\bibitem{briant:02}
A.~J. Briant, P.~Papatzacos, and J.~M. Yeomans.
\newblock {\em Philos. T. Roy. Soc. A 360}, 1792:485--495, Mar 2002.

\bibitem{japo1}
N.~Saito, Y.~Wu, K.~Hayashi, H.~Sugimura, and O.~Takai.
\newblock {\em J. Phys. Chem. B}, 107:664--667, 2003.

\end{thebibliography}

\end{document}